%% file: manuscript.tex
\documentclass[prl,aps,twocolumn,superscriptaddress]{revtex4-1}

\usepackage{graphicx} 
\usepackage{amsmath}
\usepackage{color}

\usepackage{placeins}

\usepackage{textcomp}
\usepackage[hypertexnames=false]{hyperref}

\usepackage{cleveref} %load last

\newcommand{\rcite}[1]{ref.~\cite{#1}}
\newcommand{\rcites}[1]{refs.~\cite{#1}}

\newcommand{\cohfact}{\langle \cos(\varphi) \rangle}
\newcommand{\g}{g_\mathrm{1 D}}
\newcommand{\n}{n_\mathrm{1 D}}

\newcommand*\di{\mathop{}\!\mathrm{d}}
\newcommand{\W}{W_\mathrm{eq}}

\usepackage{times}
\newcommand{\um}{\,{\textmu}m}

\begin{document} 
	
\title{Decay and recurrence of non-Gaussian correlations in a quantum many-body system}

%-----------	
\newcommand{\fu}{Dahlem Center for Complex Quantum Systems, Freie Universit{\"a}t Berlin, 14195 Berlin, Germany}
\newcommand{\tuw}{Vienna Center for Quantum Science and Technology, Atominstitut, TU Wien, Stadionallee 2, 1020 Vienna, Austria}
\newcommand{\lu}{Faculty of Mathematics and Physics,
University of Ljubljana, Jadranska 19, SI-1000 Ljubljana, Slovenia}
\newcommand{\lb}{Instituto de Telecomunica\c{c}\~{o}es, Physics of Information and Quantum Technologies Group, Av.\ Rovisco Pais 1, 1049-001, Lisbon, Portugal}
\newcommand{\lbu}{Instituto Superior T\'{e}cnico, Universidade de Lisboa, Av.\ Rovisco Pais 1, 1049-001, Lisbon, Portugal}
\newcommand{\pa}{Laboratoire Kastler Brossel, Ecole Normale Sup{\'e}rieure, Coll{\`e}ge de France, CNRS UMR 8552, Sorbonne Universit{\'e}, 24 rue Lhomond, 75005 Paris, France}

\author{Thomas Schweigler}\affiliation{\tuw}
\author{Marek Gluza}\affiliation{\fu}
\author{Mohammadamin Tajik}\affiliation{\tuw}
\author{Spyros Sotiriadis}\affiliation{\lu}
\author{Federica Cataldini}\affiliation{\tuw}
\author{Si-Cong Ji}\affiliation{\tuw}
\author{Frederik S.\ M{\o}ller}\affiliation{\tuw}
\author{Jo{\~a}o Sabino}\affiliation{\tuw}\affiliation{\lb}\affiliation{\lbu}
\author{Bernhard Rauer}\affiliation{\tuw}\affiliation{\pa}
\author{Jens Eisert}\affiliation{\fu}
\author{J{\"o}rg Schmiedmayer}\affiliation{\tuw}
%----------------

\date{\today}	
		
\maketitle 

%------- Introductory paragraph

{\bf
	Gaussian models provide an excellent effective description of many quantum many-body systems ranging from condensed matter systems \cite{Girvin,altland_simons_2010} all the way to neutron stars \cite{NeutronStars}. 
	Gaussian states are common at equilibrium when the interactions are weak.  
	Recently it was proposed that they can also emerge dynamically from a non-Gaussian initial state evolving under non-interacting dynamics~\cite{haag1973asymptotic,CramerEisertOsborne,CramerEisert,Sotiriadis_2014,GluzaEisertFarrelly,MurthySrednicki,Monnai19,Ishii19}.
	In this work, we present the experimental observation of such a dynamical emergence of Gaussian correlations in a quantum many-body system. 
	This non-equilibrium evolution is triggered by abruptly switching off the effective interaction between the observed collective degrees of freedom, while leaving the interactions between the microscopic constituents unchanged.
	Starting from highly non-Gaussian correlations, consistent with the sine-Gordon model~\cite{Schweigler17}, we observe a Gaussian state to emerge over time as revealed by the decay of the fourth and sixth order connected correlations in the quantum field.
	A description of this dynamics requires a novel mechanism for the emergence of Gaussian correlations, which is relevant for a wide class of quantum many-body systems.
	In our closed system with non-interacting effective degrees of freedom, we do not expect full thermalization~\cite{Gring12,Langen13b,Langen15,Schreiber2015,Calabrese2011,Polkovnikov2011,Eisert2015}.
	This memory of the initial state is confirmed by observing recurrences~\cite{Rauereaan7938} of non-Gaussian correlations.
}

%------- Further Introduction

Non-Gaussian states arise from the presence of interactions between the considered degrees of freedom.
Conversely, one often equates the Hamiltonian being non-interacting with the state being Gaussian.
However, this connection is not necessarily true if interactions have been present in the past.
In this case, Gaussian correlations would have to emerge dynamically from non-Gaussian states.
Such a scenario is very relevant for the emergence of statistical physics considering that the maximum entropy states of effectively non interacting systems are Gaussian.  
The results shown in this work represent the equilibration from a strongly correlated to such a Gaussian maximum entropy state.

However, despite its importance for quantum equilibration, the general phenomenon of the dynamical emergence of Gaussian correlations has only been studied theoretically, on the basis of a single type of mechanism. 
The previous theoretical work~\cite{haag1973asymptotic,CramerEisertOsborne,CramerEisert,Sotiriadis_2014,GluzaEisertFarrelly,MurthySrednicki,Monnai19,Ishii19} is concerned with the dephasing to a Gaussian state as a result of intricate spatial mixing of correlations.
This mechanism will turn out to be insufficient to explain our experimental observations. Instead, the theoretical explanation for our experimental results will rely on a different kind of correlation mixing.

%--------- introducing the model system we use 

An ideal model to study the emergence of Gaussian states is the sine-Gordon model~\cite{cuevas2014sine}, which can be quantum simulated~\cite{Cirac2012,Bloch2012} by two tunnel-coupled one-dimensional (1D) superfluids~\cite{Gritsev2007}. 
Changing the tunnel-coupling one can switch from a strongly correlated system with non-Gaussian correlations~\cite{Schweigler17} to weakly or even non-interacting effective degrees of freedom. 
The ability to change the tunneling rate, therefore, enables us to perform a sudden quench~\cite{Calabrese2006} in the interaction strength of the effective model.

%--------- experimental procedure

\begin{figure}
	\centering
	\includegraphics{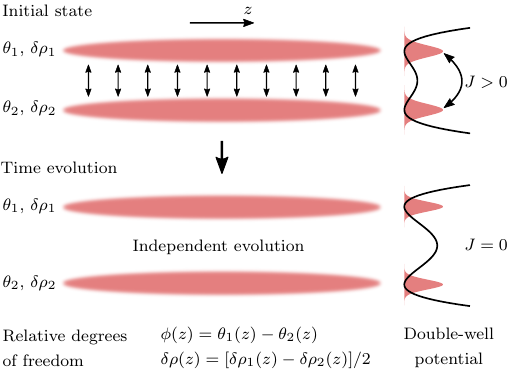}
	\caption{\textbf{Schematic of the experimental procedure.}
		We investigate the non-equilibrium dynamics of two 1D superfluids in a double-well potential.
		Each of the superfluids is described through the density fluctuations $\delta \rho_{1,2}(z)$ around their common mean density profile $\n(z)$ as well as their phase fluctuations $\theta_{1,2}(z)$. 
		From the quantities for the single superfluids, we define the relative phase fluctuations $\varphi(z)$ and the relative density fluctuations $\delta \rho(z)$ as the respective differences.
		We initially prepare the system through evaporative cooling in a double-well potential with tunneling.
		Employing a slow cooling rate, we aim at creating a thermal equilibrium state.
		Subsequently, we switch off the tunneling $J$ by ramping up the barrier separating the two wells and investigate the independent time evolution of the two superfluids.
	}
	\label{fig:schematics}
\end{figure}

In our experimental system, the superfluids are realized with ultracold bosons (\textsuperscript{87}Rb atoms) trapped in a double-well potential on an atom chip~\cite{Reichel11}.
The double-well is created through dressing with radio frequency magnetic fields~\cite{Schumm05}.
The height of the double-well barrier can be tuned by changing the amplitude of this dressing fields, thus allowing for different tunneling rates between the wells.
Using matter-wave interference, we can extract the spatially resolved phase difference $\varphi(z)$ between the superfluids.
We use a harmonic or a box-like trap for the 1D confinement of the superfluids.
A schematics of the experimental system and the quench procedure (see discussion in the following paragraphs) is depicted in \cref{fig:schematics}.

As discussed in \rcite{Schweigler17}, we can prepare states with distinctly non-Gaussian fluctuations of the relative phase $\varphi(z)$ between the superfluids through slow evaporative cooling (aiming to create a thermal equilibrium state) in the double-well potential with tunneling.
We found that the phase correlations of such states can be well described by the thermal fluctuations of the sine-Gordon model 
\begin{align}
H _ { \mathrm { SG } } = \int \mathrm{d} z \Bigg[ &\g(z) \, \delta \rho ^ { 2 } (z)  + \frac { \hbar ^ { 2 } \n(z)} { 4 m } \left( \frac { \partial  \varphi (z) } { \partial z } \right) ^ { 2 } \notag \\
&- 2 \hbar J\, \n(z) \cos \left( \varphi (z) \right) \Bigg] \label{eq:H_SG}
\end{align}
in the classical field approximation.
In addition to the relative phase fluctuations $ \varphi (z) $, the Hamiltonian also contains the conjugate field, the relative density fluctuations $\delta \rho (z)$, which are not accessible in the experiment via direct measurement.
In \cref{eq:H_SG}, $m$ represents the mass of the \textsuperscript{87}Rb atoms, $\n(z)$ the 1D mean density profile, $\g(z)$ the 1D interaction strength, and $J$ the single particle tunneling rate.
Note that \cref{eq:H_SG} is an effective model, representing a low energy approximation of the interacting microscopic Hamiltonian~\cite{Gritsev2007}.

The degree of non-Gaussianity of the state prepared by evaporative cooling depends on the phase-locking (higher for larger tunneling rate) between the superfluids.
We quantify the phase-locking strength by the coherence factor $\cohfact$.
For very weak ($\cohfact \approx 0$) and very strong phase-locking ($\cohfact \approx 1$) the fluctuations follow a Gaussian distribution, while for intermediate phase-locking we observe non-Gaussian states.
In our experimental protocol we start from such a non-Gaussian state and subsequently switch off the tunneling by ramping up the double-well barrier in approximately 2\,ms. 
Afterwards, the two superfluids evolve independently in the double-well without tunneling.
This corresponds to free evolution of the collective degrees of freedom $ \varphi (z) $ and $\delta \rho (z)$ according to \cref{eq:H_SG} with $J = 0$ (see discussion of the theoretical model below).
During the evolution, the relative phase is measured.
Since the measurement procedure is destructive, only one spatially resolved phase profile at a certain evolution time is recorded per experimental realization.

From the extracted relative phase profiles $\varphi(z)$, we calculate correlation functions using the same procedure as was used in \rcite{Schweigler17}.
Note that only phase differences between two spatial points are defined unambiguously.
Therefore, we reference the relative phase to the center of the superfluids and calculate correlation functions of $\tilde{\varphi}(z) = \varphi (z) - \varphi (0)$.
The $N$-th order equal-time phase correlation function is defined as
\begin{equation}
G^{(N)}({\boldsymbol{z}},t) = \left\langle \tilde \varphi(z_1,t) \dots \tilde \varphi(z_N,t) \right\rangle,
\label{eq:CorrelationFunction}
\end{equation}
where $\boldsymbol{z} = (z_1,\dots,z_N)$ and the expectation value is calculated by averaging over many experimental realizations.
The full correlation function, $G^{(N)}$, can be decomposed into~\cite{ZinnJustin}
\begin{equation}
G^{(N)}({\boldsymbol{z}},t) = G_{\mathrm{dis}}^{(N)}({\boldsymbol{z}},t) + G_{\mathrm{con}}^{(N)}({\boldsymbol{z}},t) .
\label{eq:DecomposeCorrFunc}
\end{equation}
The first term, $G_{\mathrm{dis}}^{(N)}$, is the {\em disconnected} part of the correlation function. 
It is fully determined by \emph{all} the lower-order correlation functions, $G^{(N')}$ with $N'<N$, and therefore does not contain new information at order $N$. 
The second term, $G_{\mathrm{con}}^{(N)}$, is the {\em connected} part of the correlation function, and contains genuine new information about the system at order $N$.
An explicit formula for $G_{\mathrm{con}}^{(N)}$ is given in \cref{general_formula_connected}. 

For Gaussian states, all higher-order correlations, $G^{(N)}$ with $N > 2$, fully factorize, i.e., all $G_{\mathrm{con}}^{(N)}$ for $N > 2$ vanish.
As a measure of non-Gaussianity of the relative phase fluctuations, we calculate the relative size of the fourth-order connected correlation function,
\begin{equation}
M^{(4)}(t)= \frac{S^{(4)}_\mathrm{con}(t)}{S_\mathrm{full}^{(4)}(t)} =\frac{\sum_{\boldsymbol{z}}{\left|G^{(4)}_{\mathrm{con}}({\boldsymbol{z}},t)\right|}}{\sum_{\boldsymbol{z}}{\left|G^{(4)}({\boldsymbol{z}},t)\right|}}. \label{eq:M4}
\end{equation}
A non-zero value of this quantity implies that the fourth-order correlation function is not simply determined by the second moments via Wick's theorem~\cite{ZinnJustin}.
Note that the experimental $z$-values lie on a discrete grid with a spacing of approximately $2${\um}, which is determined by the pixel-size of the camera used for absorption imaging.  
The sums in \cref{eq:M4} run over all distinct combinations of $z_1, \dots, z_4$ contained in a central region of the superfluids.
While we only present the results for the fourth order correlation function in the main text, we give additional plots for the sixth order correlation function in the \hyperref[sec:supp_info_6p]{supplementary information}.
Note that correlation functions of odd $N$ vanish due to symmetry.

\begin{figure}
	\centering
	\includegraphics{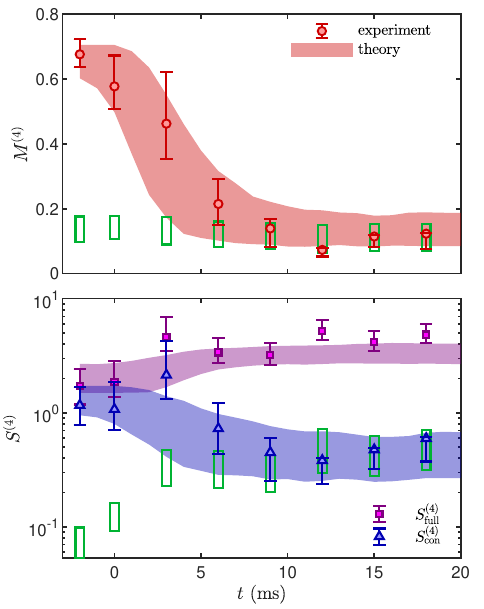}
	\caption{\textbf{Time evolution of the relative size of the fourth-order connected correlation functions.}
		The experimental results for the measure $M^{(4)}$ (upper plot, red bullets) as well as for $S_\mathrm{full}^{(4)}$ (lower plot, purple squares) and $S^{(4)}_\mathrm{con}$ (lower plot, blue triangles) are shown as a function of the evolution time, $t$.
		The first point at $t = -2$\,ms represents the initial state with a coherence factor $\cohfact_\mathrm{init} = 0.74$.
		Between $t = -2$ and $0$\,ms, the amplitude of the dressing fields is ramped up, leading to a decoupling of the two wells. 
		The shaded area represents the theory prediction for the respective quantities, considering the finite statistics and uncertainty in the decoupling time (between $-2$ and $0$\,ms), but not the uncertainty in the initial thermal coherence length, $\lambda_T$, and single particle tunneling rate $J$ (see \hyperref[sec:methods_theory]{Methods}).
		The green rectangles represent the predictions for $M^{(4)}$ (upper subplot) and $S^{(4)}_\mathrm{con}$ (lower subplot) following from Gaussian fluctuations considering the finite experimental sample size.
		All error bars as well as the vertical extension of the shaded areas and green rectangles represent 80\% confidence intervals.
		We used bootstrapping (bias-corrected and accelerated method~\cite{efron1986bootstrap}) to calculate the confidence intervals for the experimental data.
		The horizontal extension of the green rectangles was chosen arbitrarily.		
		The experimental results have been obtained for a harmonic 1D trapping potential.
		The central 50{\um} of the system are analyzed in experiment and theory, with the experimental system length being roughly 120{\um}.
	}
	% plot produced by G:\data_analysis\ComparingScanResults\2014\th_codes\non_g_codes\diffnorm_plots\thesisplots\paper_Gaussification_fig_1.m
	\label{fig:scan5100integratedevol}
\end{figure}

The experimental results for the time evolution of $M^{(4)}$ as well as $S_\mathrm{full}^{(4)}$ and $S^{(4)}_\mathrm{con}$ are presented in \cref{fig:scan5100integratedevol}.
We observe a fast decrease in $M^{(4)}$ driven by an increase in the magnitude of the phase fluctuations (increase in $S_\mathrm{full}^{(4)}$) and a decrease in the fourth-order connected correlation functions (decrease in $S^{(4)}_\mathrm{con}$).
Considering the finite experimental statistics, the resulting final state is indistinguishable from Gaussian fluctuations.
The Gaussian model that we compare with in the figures is simply given by the experimental mean values $\langle \tilde\varphi (z,t) \rangle$ and covariance matrix $\langle \tilde\varphi (z,t) \tilde\varphi (z^\prime,t) \rangle$ at the particular evolution time $t$. 
Note that the increase in the overall fluctuations corresponds to a decrease in the integrated interference contrast as previously observed~\cite{Gring12,Langen13b,Rauereaan7938} and can be explained by dephasing of non-interacting eigenmodes~\cite{Kitagawa2011,Langen18}.
Additional measurements for different initial phase-locking strengths show similar behavior (see \cref{fig:suppinfoadditionalresults}).

Let us now discuss our theoretical model in more detail.
As already mentioned above, we know that the relative phase fluctuations of the initial state can be well described with the sine-Gordon model \eqref{eq:H_SG} in thermal equilibrium and classical fields approximation~\cite{Beck18}.
We use this description for both the initial relative density and phase fluctuations.
Our model for the initial state, therefore, leads to non-Gaussian phase fluctuations whose magnitude gets suppressed with increased tunneling.
The non-Gaussianity is a direct consequence of the beyond quadratic tunneling term in $H _ { \mathrm { SG } }$ (last term in \cref{eq:H_SG}). 
The suppression with increased tunneling can be understood from an energy argument and is explicitly discussed in the \hyperref[sec:supp_info_rel_size]{supplementary information}.
Meanwhile, the density fluctuations are Gaussian and independent of the tunneling rate since there is only a quadratic, $J$ independent term of $\delta \rho$ in $H _ { \mathrm { SG } }$.
Cross-correlations between density and phase fluctuations vanish since there are no cross terms in \cref{eq:H_SG}.

For the time evolution we simply set $J$ in \cref{eq:H_SG} to zero giving the Luttinger liquid model.
As already discussed, this is a non-interacting model for the collective degrees of freedom, the relative phase and density fluctuations. 
The quadratic effective Hamiltonian can be diagonalized in terms of its eigenmodes.
We get
\begin{equation}
\tilde\varphi_n(t) = \tilde\varphi_n (0) \cos \left(\epsilon_n t / \hbar \right) - C_n \, \delta \tilde\rho_n (0) \, \sin \left(\epsilon_n t / \hbar \right)
\label{eq:phonon_t_evo}
\end{equation}
for the eigenmode expansion of the phase ($\tilde\varphi_n$) and density ($\delta \tilde\rho_n$) fluctuations.
Here, we denote the eigenenergies by $\epsilon_n$ and $C_n$ is a constant depending on the mode number $n$.
From \cref{eq:phonon_t_evo}, one can intuitively understand the experimental results.
The large Gaussian density fluctuations of the initial state rotate into the phase quadrature while the small initial non-Gaussian phase fluctuations (suppressed by the tunneling) rotate out.
Focusing only on a single eigenmode, this rotation would lead to an oscillatory behavior for the degree of non-Gaussianity.
However, several eigenmodes contribute to a typical local or global observable.
Dephasing between the contributing modes leads to a suppression of the oscillatory behavior.
Together, rotation and dephasing result in a fast decrease of the relative size $M^{(4)}$ of the fourth-order connected correlation function.

In other words, the emergence of the Gaussian phase correlations relies on the mixing between the initially suppressed non-Gaussian phase fluctuations and the non-suppressed initial Gaussian density fluctuations. 
This is in contrast to former theoretical work relying on spatial mixing of the correlations~\cite{haag1973asymptotic,CramerEisertOsborne,CramerEisert,Sotiriadis_2014,GluzaEisertFarrelly,MurthySrednicki,Monnai19,Ishii19}.
Since in our case the dynamics is not delocalizing (see \hyperref[sec:supp_info_ps_mix]{supplementary information}), spatial mixing cannot explain the experimental results.
Instead,
the initial Gaussian density fluctuations are crucial for observing the dynamical emergence of Gaussian phase correlations in our model.
In theory calculations with initial density fluctuations set to zero, a statistically significant fourth-order connected part remains at all times (see \hyperref[sec:supp_info_dens_import]{supplementary information}).

The 1D mean density profile $\n$, and, as a consequence, also the 1D interaction strength $\g$ (see \cref{eq:g1D_broadened}) are position ($z$) dependent for a non-homogeneous system.
The form of the eigenmodes, the eigenenergies $\epsilon_n$ and the constant $C_n$ in \cref{eq:phonon_t_evo} depend on the form of this spatial dependencies (determined by the trapping geometry) and the choice of boundary conditions.
A harmonic trapping potential leads to an incommensurate set of eigenenergies~\cite{Geiger_2014}.
The different eigenmodes dephase with respect to each other leading to an apparent equilibration with the initial phase (density) fluctuations being distributed to both the phase and density quadratures~\cite{Kitagawa2011,Gring12}.
No rephasing is expected and the phase correlations should remain Gaussian after the initial dephasing.
It appears like a maximum entropy state has been reached~\cite{Langen15}.

\begin{figure*}
	\centering
	\includegraphics{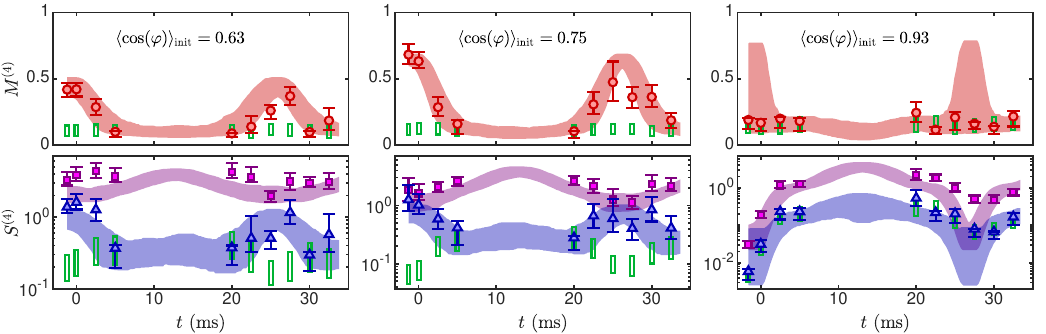}
	\caption{\textbf{Recurrence of the non-Gaussian phase fluctuations.}
		As for \cref{fig:scan5100integratedevol} (see there for the meaning of the plotted quantities and error bars), but for a 50{\um} long near-homogeneous system (box-like 1D confinement) and several different initial phase-locking strengths quantified by $\cohfact_\mathrm{init}$.
		In all cases a recurrence of coherence~\cite{Rauereaan7938} can be observed around 25\,ms, indicated by a dip in the value of $S_\mathrm{full}^{(4)}$.
		In the measurements with $\cohfact_\mathrm{init} = 0.63$ or 0.75, the initial state exhibits non Gaussian phase correlations, and we also observe a recurrence in the relative ($M^{(4)}$) and absolute ($S^{(4)}_\mathrm{con}$) size of the connected fourth-order correlation function.
		In the measurement with $\cohfact_\mathrm{init} = 0.93$, the initial phase fluctuations are Gaussian, and no such peak in $M^{(4)}$ occurs.
		This supports the picture that the recurrences of non-Gaussian fluctuations are due to memory of the initial state being preserved.
		The central 38{\um} of the system are analyzed in experiment and theory.
		To get better statistics for the presented times we decided not to measure at intermediate times between the initial dephasing and the recurrence.
	}		
	% plot produced by G:\data_analysis\ComparingScanResults\2014\th_codes\non_g_codes\diffnorm_plots\thesisplots\paper_Gaussification_recurrence_different_scales.m
	\label{fig:recurrences}
\end{figure*}

For a homogeneous system on the other hand, the eigenenergies are equally spaced, the eigenmodes are given by sine and/or cosine functions depending on the boundary conditions.
After the initial dephasing of the eigenmodes, we expect a rephasing at a later time with the system returning to its initial state.  
Indeed, we observed a recurrence of phase coherence in \rcite{Rauereaan7938} where a box-like confinement leading to a fairly homogeneous system was used.
Using the same kind of confinement, but starting from a strongly non-Gaussian initial state, we now also observe a recurrence of non-Gaussian phase fluctuations as presented in \cref{fig:recurrences}.
At the recurrence time, the dominant Gaussian fluctuations get rotated back---simultaneously for all eigenmodes---into the density fluctuation sector and the non-Gaussianity reappears.

The experimental results presented in \cref{fig:scan5100integratedevol,fig:recurrences} are in good agreement with the predictions of our theoretical model (see \hyperref[sec:methods_theory]{Methods} for details of the calculations).
Note that the evolution times presented in \cref{fig:scan5100integratedevol}  are rather short compared to the system size divided by the speed of sound ($\approx 2${\um}/ms).
Therefore, the exact trapping geometry does not matter too much as long as we analyze the approximately homogeneous central part of the system only.
For simplicity, we therefore perform the theory calculations presented in \cref{fig:scan5100integratedevol} for a large homogeneous system in order to avoid any boundary effects.
Meanwhile, the calculations shown in \cref{fig:recurrences} are done for a homogeneous system of the same length as the experimental box-like confinement and Neumann boundary conditions for the phase $\varphi$.
This represents an ideal hard-walled box.
Here, the boundary and finite size effects are crucial for correctly describing the recurrences.

While our experiment has uncovered a mechanism for the dynamical emergence of Gaussian correlations distinct from the previously studied spatial mixing phenomena~\cite{haag1973asymptotic,CramerEisertOsborne,CramerEisert,Sotiriadis_2014,GluzaEisertFarrelly,MurthySrednicki,Monnai19,Ishii19} it is instructive to also stress their similarities: 
In both cases there is initially a Gaussian component of the system's state that plays the role of a Gaussian `bath' into which the non-Gaussian component gets diluted by dephasing. 
In our case the Gaussian bath is clearly separated in the density sector, while in the case of spatial mixing it can be identified as the prevalence of Gaussian correlations at distant points due to clustering.

For our experiment, before decoupling, the phase fluctuations enter with a large weight into the definition of the creation and annihilation operators for the eigenmodes diagonalizing the effective Hamiltonian.
At the moment when the tunneling is switched off, the effective Hamiltonian changes and the relative weight of the phase fluctuations gets reduced with respect to the weight of the density fluctuations.	
While the correlations are not changed during the quench and no additional Gaussian fluctuations are produced, the reduced importance of the non-Gaussian phase fluctuations compared to the Gaussian density fluctuations can be interpreted as the creation of the Gaussian bath.

In addition to the dephasing of the phononic modes discussed above, further equilibration might occur in our system due to additional terms beyond second order not considered in the low energy effective model stated in \cref{eq:H_SG}.
Such terms lead to the damping of the recurrence heights observed in \rcite{Rauereaan7938}.
We expect that the additional terms also lead to a damping of the recurrences of $M^{(4)}$ when investigating longer evolution times or bigger sample sizes.
The investigation of such a damping will be the objective of future studies.
Paradoxically, it is typically such weak non-quadratic terms that bring about irreversibility in the dynamics towards the final approximately Gaussian steady state,
similarly to the classical statistical mechanics of the ideal gas, where infinitesimal scattering is necessary to reach thermal equilibrium.

Finally, let us discuss the broad applicability of our new mechanism for the emergence of Gaussian correlations beyond the specifics of the actual system considered.
Firstly, note that the sine-Gordon model is a special case of a generic scalar field theory in 1D described by the Hamiltonian
\begin{equation}
H=\int{ \mathrm{d} z \left[a \, \pi(z)^2+ b\, \left( \frac { \partial  \phi (z) } { \partial z } \right)^2+V\left(\phi(z)\right)\right]}.\label{eq:general_Hamiltonian}
\end{equation}
Here we denote the canonically conjugate fields by $\phi$ and $\pi$ following the standard notation found in field theory textbooks.	
\Cref{eq:general_Hamiltonian} represents the general form of a renormalizable scalar field theory~\cite{srednicki2007quantum} and is of great importance for the effective description of a large variety of quantum many-body systems~\cite{ZinnJustin}.

All the ingredients necessary for the mechanism explaining our experimental results with the sine-Gordon theory are also present for the more general Hamiltonian \cref{eq:general_Hamiltonian} when performing the same kind of quench procedure.
I.e., we start with a thermal equilibrium state in the high temperature limit with a beyond quadratic $V(\phi)$ and subsequently set $V(\phi)$ to zero. 
During the following evolution, the previously strongly-correlated $\phi$ sector has an influx of thermal Gaussian $\pi$ fluctuations, which dominate because, unlike those of $\phi$, they were not suppressed by the energetic penalty $V(\phi)$ (see the \hyperref[sec:supp_info_rel_size]{supplementary information} for a more formal argument). 
This is exactly the generalization of what has been discussed above for the specific case of the sine-Gordon model.
Moreover, we conjecture that our mechanism is not restricted to theories of the form of \cref{eq:general_Hamiltonian}, but also present in other theoretical models where non-Gaussian and Gaussian fluctuations are present in different sectors.
However, to identify the full range of applicability of our new mechanism goes beyond the scope of the present work.

To conclude, we uncover a pathway for the dynamical emergence of Gaussian correlations. This phenomenon has received too little attention so far, considering its paramount importance in the wider context of quantum equilibration and the emergence of statistical mechanics from an underlying microscopic quantum evolution. On a broader level, our work highlights the power of precise and quantitative experimental studies to unveil the foundations of statistical mechanics at the border to the quantum world.

\section{Methods}
\label{sec:methods}

\subsection{Details of the experimental realization}
\label{sec:methods_exp}

The (tunnel-coupled) 1D superfluids are realized using ultracold gases of \textsuperscript{87}Rb in a double-well potential on an atom chip~\cite{Reichel11}. 
Each well consists of a highly elongated cigar-shaped trap (two tightly confined directions, one direction with weak confinement, corresponding to the 1D direction $z$). 
The wells are separated along one of the tightly confined directions (see \cref{fig:schematics}). 
The separation is horizontal, avoiding the influence of gravity. 
By tuning the height of the barrier separating the wells we can change the tunnel-coupling between the two superfluids. 

The clouds are initially prepared by evaporatively cooling the atoms whilst keeping the double-well trap static. 
The relative evaporation rates at the end of the cooling ramp amount to a few percent per 10\,ms.
Right after the end of the evaporative cooling, the double-well barrier is ramped up to start the non-equilibrium evolution (see discussion in the main text and \cref{fig:schematics}).
We do not expect any substantial heating due to technical noise during the non-equilibrium evolution~\cite{Rauereaan7938,kuhnert2013thesis}.

The measured harmonic frequency for the weakly confined direction is $\omega_z \approx 2\pi\times7\,$Hz, leading to clouds of approximately 120{\um} length. 
In addition to this harmonic magnetic confinement we superimpose a box-shaped optical potential for some of the measurements.
For the results presented in \cref{fig:recurrences}, the optical potential was created by a laser with 660\,nm wavelength (blue detuned) and shaped by a digital micromirror device (DMD)~\cite{Tajik:19}.
For the measurements presented in \cref{fig:suppinfoadditionalresults}, a 767\,nm (blue detuned) laser was used together with a simple mask~\cite{Rauereaan7938,rauer2018thesis}.

\subsection{Calculation of the correlation functions and their connected parts}

As defined in \cref{eq:CorrelationFunction} in the main text, we evaluate the phase correlation functions as
\begin{equation}
G^{(N)}({\boldsymbol{z}},t) = \left\langle[\varphi(z_1,t)- \varphi(0,t)]\dots[\varphi(z_N,t) - \varphi(0,t)]\right\rangle,
\end{equation}
where the brackets denote the averaging over different experimental realizations and $\varphi(z)$ is extracted via matter-wave interference. 
Similarly, we calculate the connected part using~\cite{shiryaev2016probability}
\begin{align}
\begin{split}
G_{\mathrm{con}}^{(N)}({\boldsymbol{z}},t) =  \sum_{\pi} \Bigg[ \ &(|\pi|-1)! \ (-1)^{|\pi|-1}  
\\ &\prod_{B \in \pi} \left\langle \prod_{i \in B} [\varphi(z_i,t)- \varphi(0,t)] \right\rangle \Bigg]  . \label{general_formula_connected}
\end{split}
\end{align}
Here the sum runs over all possible partitions $\pi$ of $\{1,\dots,N\}$, the first (left) product runs over all blocks $B$ of the partition and the second (right) product runs over all elements $i$ of the block. 
$|\pi|$ is the number of blocks in the partition. 
Note that \cref{general_formula_connected} does not represent an unbiased estimator of the connected correlation function. 
However, for our large sample sizes the bias should be negligible.

\subsection{Details of the theory calculation}	
\label{sec:methods_theory}

The theoretical calculations are performed numerically.
Moreover, a homogeneous system is assumed.
For sampling the phase fluctuations of the initial state (thermal fluctuations of the sine-Gordon model), we use the method discussed in \rcites{Beck18,Schweigler17}.
This method produces numerical realizations of a certain length, which represents a part of an infinite system.
The thermal fluctuations for the sine-Gordon model in classical fields approximation are determined by the thermal coherence length, $\lambda_T$, and the single particle tunneling rate, $J$~\cite{Schweigler17,Grisins2013}.
Self consistently fitting both $\lambda_T$ and $J$ from the initial state is very challenging since the parameters are correlated for the attempted fitting procedures~\cite{Schweigler19thesis}.
In order to avoid overfitting or large fitting errors, we use $\lambda_T = 11 ${\um} as a reasonable value for all theoretical calculations presented in the paper and supplementary information.
Having set the value of $\lambda_T$, we self consistently fit $J$ from the initial values of $\cohfact$ leading to the different values of $J$ used in the theory calculations for the different measurements. 
The experimental value of $\cohfact_\mathrm{init}$ used for the fit is obtained by averaging over the experimental realizations as well as the same central interval of the superfluids used also for all other analysis.
Note that both the evolution of $M^{(4)}$ and $S_\mathrm{con}^{(4)}$ is rather insensitive to the value of $\lambda_T$.
In other words, the conclusions concerning the dynamical emergence of Gaussian correlations and the recurrence of non-Gaussian correlations are insensitive to the particular choice of $\lambda_T$.
In contrast, the evolution of $S_\mathrm{full}^{(4)}$ strongly depends on $\lambda_T$.
Given our choice of $\lambda_T$, the experimental results and theory predictions for $S_\mathrm{full}^{(4)}$ agree fairly well in most cases.

Numerical sampling of the thermal density fluctuations is straightforward as the fluctuations are Gaussian.
The thermal density fluctuations for the sine-Gordon model in classical fields approximation are determined by the 1D interaction strength, $\g$, and again by the thermal coherence length, $\lambda_T$.
We use the density broadened 1D interaction strength~\cite{Rauereaan7938,rauer2018thesis,Schweigler19thesis}
\begin{equation}
\g = \hbar \omega_{\perp} a_{\mathrm{s}} \frac{2+3 a_{\mathrm{s}} \n}{\left(1+2 a_{\mathrm{s}} \n\right)^{3 / 2}},
\label{eq:g1D_broadened}
\end{equation}
for both the initial state and the time evolution.
In \cref{eq:g1D_broadened}, $a_{\mathrm{s}} = 5.2$\,nm \cite{van_Kempen_2002} is the 3D scattering length.
The transverse trapping frequency is $\omega_{\perp} = 2 \pi \times 1.4$\,kHz for the results presented in \cref{fig:scan5100integratedevol} and the supplementary information.
For \cref{fig:recurrences} on the other hand, we use $\omega_{\perp} = 2 \pi \times 1.45$\,kHz.
For the homogeneous 1D atomic density, $\n$, we use the value averaged over the analyzed central part of the superfluids.

Having obtained numerical realizations for the initial state, we numerically evolve them with the discretized version of \cref{eq:H_SG} and $J = 0$.
We use Neumann boundary conditions and a homogeneous system, i.e., a constant $\n$ and $\g$.
For the results presented in \cref{fig:scan5100integratedevol} and the supplementary information, we use a theoretical system size of $L = 200${\um}.
Therefore, as desired, we do not get any influence of the boundaries for the investigated time scales in the central part of the system.
The outcome mimics the results for an infinite system.
For the theory predictions presented in \cref{fig:recurrences}, we use the actual length $L = 50${\um} of the experimental system.

In order to correctly consider the effect of the finite experimental sample size, we set the number of numerical realizations equal to the experimental sample size of the corresponding measurements.
The experimental sample sizes vary between 118 and 623 (median 390) for the different measurements.
The calculations are then repeated 200 times to obtain the confidence intervals presented in the figures.

For all presented theory predictions, the effect of the finite experimental spatial resolution is approximated by convolving the numerically calculated phase profiles with a Gaussian of a standard deviation of $\sigma_\mathrm{PSF} = 3${\um} \cite{Schweigler19thesis} before calculating the presented quantities.

\section*{Acknowledgments} 
We thank Igor Mazets, Sebastian Erne, Thomas Gasenzer, J{\"u}rgen Berges and Tim Langen for helpful discussions.
This work is supported by the DFG/FWF Collaborative Research Centre ``SFB 1225 (ISOQUANT)'', and  the ESQ Discovery Grant ``Emergence of physical laws: From mathematical foundations to applications in many body physics'' of the Austrian Academy of Sciences (ÖAW).
F.C., F.M., B.R., J.Sa., and T.S.\ acknowledge support by the Austrian Science Fund (FWF) in the framework of the Doctoral School Complex Quantum Systems (CoQuS).
S.S.\ acknowledges support by the Slovenian Research Agency (ARRS) under grant QTE (N1-0109) and by the ERC Advanced Grant OMNES (694544).
S.J.\ acknowledges supported by an Erwin Schr{\"o}dinger Quantum Science \& Technology (ESQ) Fellowship funded through the European Union's Horizon 2020 research and innovation programme under the Marie Sklodowska-Curie grant agreement No.\ 801110.
J.Sa.\ acknowledges support by the Funda\c{c}{\~a}o para a Ci{\^e}ncia e a Tecnologia (PD/BD/128641/2017).
J.E.\ acknowledges funding from the DFG (FOR 2724, EI 519/9-1, EI 519/7-1, CRC 183), the FQXi, and the European Union's Horizon 2020 research and innovation programme under grant agreement No.\ 817482 (PASQuanS).
J.E., M.G., J.Sch., and S.S.\ would like to thank the Erwin Schr{\"o}dinger Institute for its hospitality and support under the programme ``Quantum Simulation - from Theory to Application'' (LCW 2019).

\section*{Author contributions}
T.S.\ performed the experiment and data analysis with contributions by M.T., B.R., F.C., S.J., F.M., and J.Sa.
T.S.\ did the theory calculations with contributions by M.G.\ and S.S.
J.Sch.\ and J.E.\ provided scientific guidance in experimental and theoretical questions.
J.Sch.\ conceived the experiment.
All authors contributed to the interpretation of the data and to the writing of the manuscript.
	
\section*{Data availability}
The experimental raw data (absorption images) for \cref{fig:scan5100integratedevol,fig:recurrences} can be found in~\rcite{rawdata_Gaussification}.
%The numerical values of the quantities plotted in \cref{fig:scan5100integratedevol,fig:recurrences}, and Extended Data \cref{fig:suppinfoadditionalresults} are provided as a Supplementary Information files.
All other data are available from the authors upon reasonable request.

\bibliography{biblio}	

%-----Extended Data	

\setcounter{figure}{0}
\renewcommand{\thefigure}{E\arabic{figure}}

\FloatBarrier
\pagebreak		
	
\begin{figure*}
	\centering
	\includegraphics{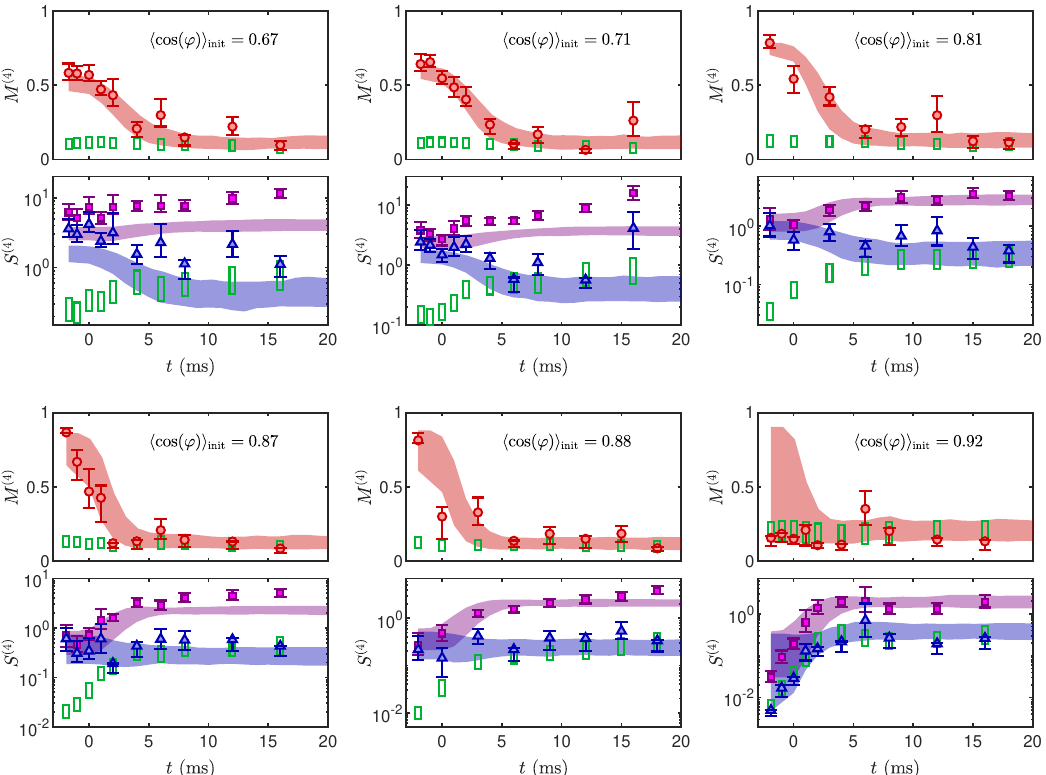}
	\caption{
		\textbf{Additional results for the time evolution of the relative size of the fourth-order connected correlation functions.}
		As for \cref{fig:scan5100integratedevol} (see there for the meaning of the plotted quantities and error bars), but for several different initial phase-locking strengths (quantified by $\cohfact_\mathrm{init}$) and different trapping geometries.
		The results for a $\cohfact_\mathrm{init}$ of 0.81 and 0.88 have been obtained with a harmonic confinement. 
		For all other results, a 75{\um} long box trap has been superimposed onto the harmonic confinement (see \hyperref[sec:methods_exp]{Methods}).
		We find that Gaussian correlations emerge dynamically for all measurements, independent of the initial phase-locking strength or the trapping geometry.
		The speed for the decay of $M^{(4)}$ increases with the initial phase-locking in agreement with our theoretical model.
		One can understand the trend by realizing that the phase-fluctuations of the initial state get smaller with increasing phase-locking and are therefore more quickly overshadowed by the mixed-in initial density fluctuations.	
	}
	\label{fig:suppinfoadditionalresults}
	% plot produced by G:\data_analysis\ComparingScanResults\2014\th_codes\non_g_codes\diffnorm_plots\thesisplots\paper_Gaussification_supp_info_plot.m
\end{figure*}

%---- supp. info
	
\appendix		
\widetext
	
\FloatBarrier		
\pagebreak

\input{supp_info}

\end{document}

%% file: supp_info.tex
\setcounter{equation}{0}
\setcounter{figure}{0}

\makeatletter
\renewcommand{\theequation}{S\arabic{equation}}
\renewcommand{\thefigure}{S\arabic{figure}}

\FloatBarrier

\begin{center}
	\textbf{\large Supplementary Information}
\end{center}

\section{Results for the sixth-order correlation functions}
\label{sec:supp_info_6p}

\begin{figure}
	\centering
	\includegraphics{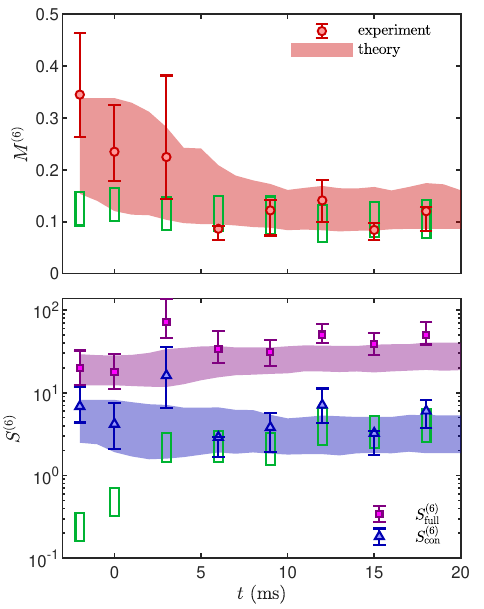}
	\caption{\textbf{Time evolution of the relative size of the sixth-order connected correlation functions.}
		As for \cref{fig:scan5100integratedevol} (see there for the meaning of the plotted quantities and error bars), but for the sixth-order instead of the fourth-order correlation functions.
	}
	% plot produced by G:\data_analysis\ComparingScanResults\2014\th_codes\non_g_codes\diffnorm_plots\thesisplots\paper_Gaussification_fig_1.m
	\label{fig:scan5100integratedevol6p}
\end{figure}

\begin{figure}
	\centering
	\includegraphics{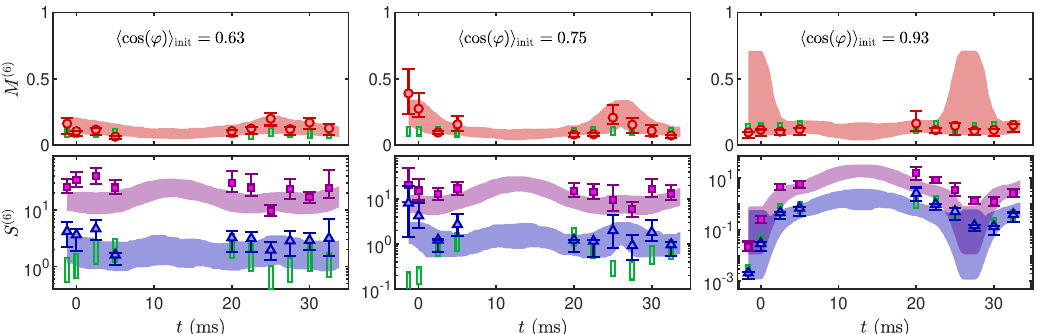}
	\caption{\textbf{Recurrence of the sixth-order correlation functions.}
		As for \cref{fig:recurrences}, but for the sixth-order instead of the fourth-order correlation functions.
		See the caption of \cref{fig:scan5100integratedevol} for the meaning of the plotted quantities and error bars.
	}
	% plot produced by G:\data_analysis\ComparingScanResults\2014\th_codes\non_g_codes\diffnorm_plots\thesisplots\paper_Gaussification_recurrence_different_scales.m
	\label{fig:recurrencesdifferentscales6p}
\end{figure}

\begin{figure}
	\centering
	\includegraphics{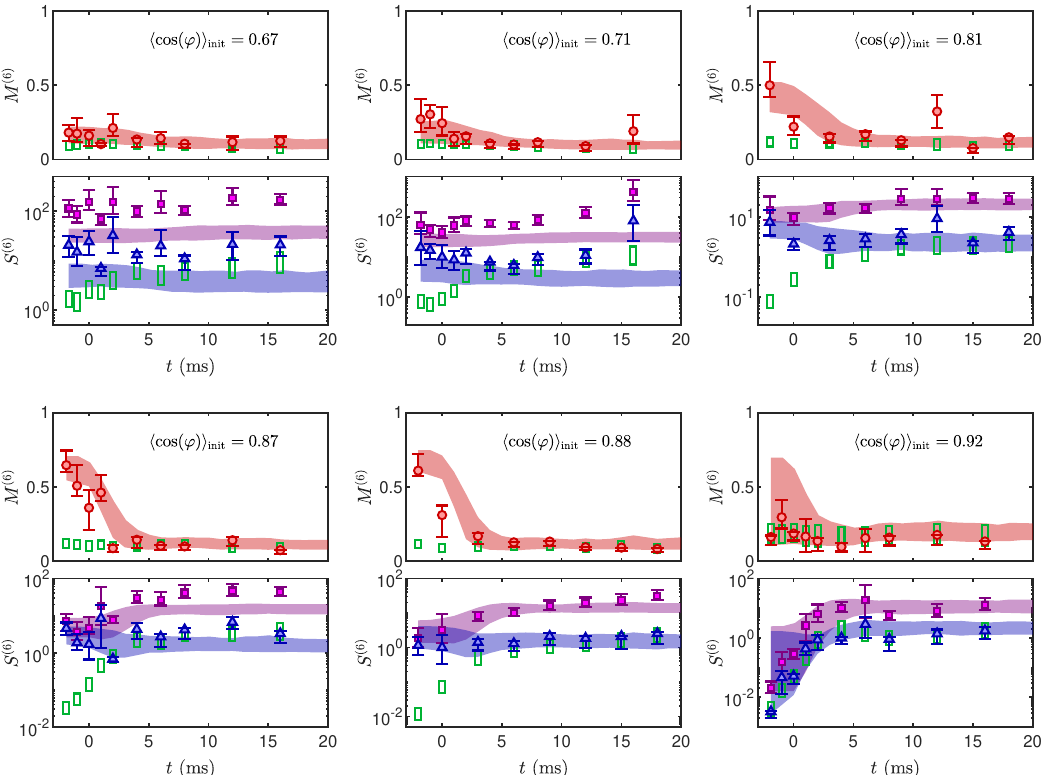}
	\caption{\textbf{Time evolution of the relative size of the sixth-order connected correlation functions.}
	As for \cref{fig:suppinfoadditionalresults}, but for the sixth-order instead of the fourth-order correlation functions.
	See the caption of \cref{fig:scan5100integratedevol} for the meaning of the plotted quantities and error bars.
	}
	% plot produced by G:\data_analysis\ComparingScanResults\2014\th_codes\non_g_codes\diffnorm_plots\thesisplots\paper_Gaussification_supp_info_plot.m
	\label{fig:suppinfoadditionalresults6p}
\end{figure}

The results for the sixth-order correlation functions corresponding to the fourth-order results presented in \cref{fig:scan5100integratedevol}, \ref{fig:recurrences} and \ref{fig:suppinfoadditionalresults} are shown in \cref{fig:scan5100integratedevol6p,fig:recurrencesdifferentscales6p,fig:suppinfoadditionalresults6p}.
In analogy to \cref{eq:M4} we have
\begin{equation}
M^{(6)}(t)= \frac{S^{(6)}_\mathrm{con}(t)}{S_\mathrm{full}^{(6)}(t)} =\frac{\sum_{\boldsymbol{z}}{\left|G^{(6)}_{\mathrm{con}}({\boldsymbol{z}},t)\right|}}{\sum_{\boldsymbol{z}}{\left|G^{(6)}({\boldsymbol{z}},t)\right|}}. \label{eq:MN}
\end{equation}
We get qualitatively similar results for the sixth-order as for the fourth-order correlation functions.
However due to the more challenging requirements concerning statistics and systematic errors we get bigger error bars and slightly worse agreement between experiment and theory for the sixth order.

\section{Additional discussion of the theoretical model}

\paragraph{Importance of the initial density fluctuations}	
\phantomsection
\label{sec:supp_info_dens_import}	
	
\begin{figure}
	\centering
	\includegraphics{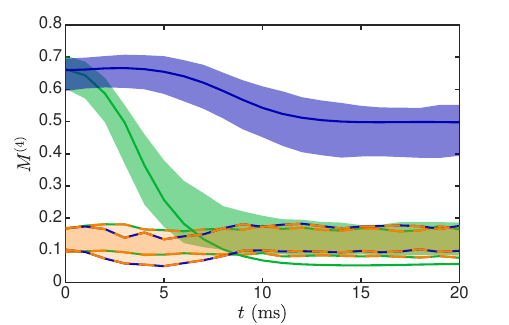}
	\caption{\textbf{Theory predictions with (green) and without (blue) initial density fluctuations.}
		For the parameters of the measurement shown in \cref{fig:scan5100integratedevol}.
		The solid lines give the results for $10^5$ numerical realizations while the shaded areas give the uncertainty considering the experimental sample size of 378.
		The orange shaded areas bounded by the dashed lines represent the predictions for Gaussian fluctuations. 
		The green-orange and blue-orange dashed boundary lines mark the results for the case with and without initial density fluctuations respectively. 
		As one can see, with initial density fluctuations, the final state appears Gaussian (assuming the experimental sample size) while the theory prediction without initial density fluctuations stays distinctively non-Gaussian.
	}
	\label{fig:scan5100theowwodens}
	% plot produced by G:\data_analysis\ComparingScanResults\2014\th_codes\non_g_codes\diffnorm_plots\thesisplots\paper_Gaussification_theo_w_wo_dens.m
\end{figure}

As already discussed in the main text, in our theoretical model, the dynamical emergence of Gaussian correlations is due to the rotating of large Gaussian fluctuations from the density to the phase quadrature. 
In order to illustrate this explicitly, we compare the theory predictions for the usual case with initial density fluctuations and the fictitious case of having no initial density fluctuations.
As can be seen from \cref{fig:scan5100theowwodens}, a significant connected fourth-order correlation remains at all times for the theory calculation without initial density fluctuations, consistently with general theoretical predictions for Luttinger liquid dynamics~\cite{Sotiriadis_LL,Sotiriadis_2017}. 
This observation demonstrates the crucial importance of the Gaussian initial density fluctuations for the dynamical emergence of Gaussian correlations, in our model.
Note that also without initial density fluctuations the theory predicts a slight decrease in $M^{(4)}$. 
This decrease is due to the mixing of correlation from different spatial points, as discussed further below.	

\paragraph{Relative size of the initial phase and density fluctuations}
\phantomsection
\label{sec:supp_info_rel_size}

\begin{figure}
	\centering
	\includegraphics{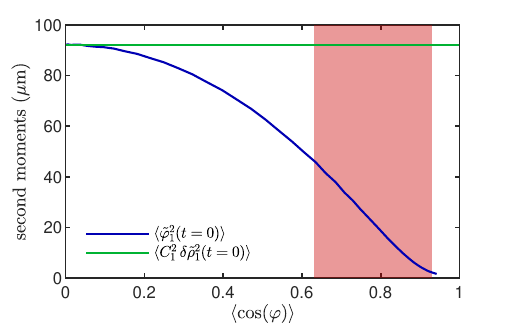}
	\caption{\textbf{Magnitude of the initial phase and density fluctuations in the theoretical model.}
	Cosine transformed initial phase (blue line) and density (green line) fluctuations. 
	The results for $k_1 = \pi/L$ are shown as a function of the phase locking strength.
	The system size $L = 50${\um} is chosen to be consistent with \cref{fig:recurrences}.
	The results for the phase fluctuations are obtained numerically (see \hyperref[sec:methods_theory]{Methods}) while for the density fluctuations $\langle C_n^2 \, \delta \tilde\rho_n^2 (t=0)\rangle = 4/(\lambda_T \, k_n^2)$ is used.
	As for all other presented theory calculations $\lambda_T = 11${\um}.
	The range of phase-locking used in the experiment is indicated by the red shaded area.
	The finite experimental spatial resolution is considered for $\cohfact$ but not for the second moments. 
	}
	\label{fig:initialstatefluctuations}
	% plot produced by G:\data_analysis\ComparingScanResults\2014\th_codes\onePI\Gaussification_initial_state.m
\end{figure}

Let us now explicitly show that in our theoretical model the initial phase fluctuations are suppressed compared to the initial density fluctuations. 
For the comparison of the two quantities, we have to consider the factor $C_n$ from \cref{eq:phonon_t_evo}, i.e., we want to compare $\langle\tilde\varphi_n^2 (t=0) \rangle$ with $ \langle C_n^2 \, \delta \tilde\rho_n^2 (t=0)\rangle$.
Remember that $\tilde\varphi_n$ and $\delta \tilde\rho_n$ represent the eigenmode expansions of the phase and density fluctuations respectively.
Assuming a homogeneous system with Neumann boundary conditions, the eigenmodes are given by cosine functions.
We get
\begin{equation}
	C_n = \frac{2\g}{\hbar c k_n},
\end{equation} 
where $k_n$ is the wavenumber taking the values
\begin{equation}
	k_n = n \frac{\pi}{L}
\end{equation}
with $n$ being a positive integer. 
The speed of sound is denoted by $c$ and can be calculated as
\begin{equation}
	c = \sqrt{\frac{\g \n}{m}}.
\end{equation}
In \cref{fig:initialstatefluctuations} the results for the lowest non-zero mode ($k_1 = \pi/L$) are shown as a function of the phase-locking.
We see that for the parameter range used in the experiment, the phase fluctuations are suppressed compared to the density fluctuations.
For higher $k_n$ this suppressing will get less.
However, the integral quantities in \cref{eq:M4} will be dominated by the lowest lying modes.

More generally, connecting to the discussion in the main text, we can argue that any viable potential $V(\phi(z))$ in Hamiltonians of the form
\begin{equation}
H=\int{ \mathrm{d} z \left[a \, \pi(z)^2+ b\, \left( \frac { \partial  \phi (z) } { \partial z } \right)^2+V\left(\phi(z)\right)\right]}
\end{equation}
will suppress the fluctuations in $\phi$, at least in the classical fields approximation.
This can be seen from the formalism introduced in \rcite{Beck18}. 
There the $\phi$ fluctuations are calculated in the classical fields approximation by solving the stochastic differential It\={o} equation
\begin{equation} 
\mathrm{d}\phi = A(\phi) \di z + \sqrt{2D}\, \di X.
\label{Ito} 
\end{equation} 
Here $\mathrm{d}\phi$ describes the stochastic change in the field when moving $\di z$ in the spatial 1D direction.
The change  $\di \phi$ is determined by the deterministic drift term $A(\phi) \di z$ and the stochastic diffusion term $\sqrt{2D}\, \di X$,
where $\di X$ is an infinitesimally small and uncorrelated random term obeying
Gaussian statistics with zero mean and the variance equal to $\di z$.
The stochastic realizations for $\phi(z)$ are then obtained by sampling the initial values (say, at $z=0$) of the field from its equilibrium distribution $\W (\phi)$ and subsequently integrating \cref{Ito}.

The drift coefficient is connected to the equilibrium distribution via
\begin{equation}
	A(\phi) = D \frac {\di \ln W_\mathrm{eq} (\phi)}{\di \phi}.
\end{equation}
It depends on the potential $V(\phi)$ since $W_\mathrm{eq} (\phi)$ depends on it.
We now want to check whether starting from a certain value of $\phi$, the drift coefficient $A(\phi)$ leads to a force against or in support of the diffusion.
Therefore, we simple have to calculate the derivative of $A(\phi)$.
If the derivative is negative, then we have a reversing force.
To know whether the drift decreases fluctuation in general, we simply calculate the average
\begin{equation}
	\int \di \phi \, \W (\phi) \frac {\di A(\phi)}{\di \phi}  = D \int \di \phi \, \W \frac {\di^2 \ln \W (\phi)}{\di \phi^2} = D \left[ \int \di \phi \, \frac {\di^2 \W (\phi)}{\di \phi^2} - \int \di \phi \, \frac{1}{\W} \left(\frac {\di \W (\phi)}{\di \phi} \right)^2 \right]. \label{eq:av_back_force}
\end{equation} 
The first integral on the right hand side is just a surface term and vanishes with proper assumptions for $\W$.
In order to be normalizable the equilibrium distribution needs to fulfill $\lim_{\phi \to \pm \infty} \W = 0$ which leads to a vanishing surface term for the integral limits $\pm \infty$.
However, not only normalizable, but also periodic $\W$ can occur like in the case of the sine-Gordon model.
In case of a periodic $\W$ the integration in \cref{eq:av_back_force} is performed over one period.
The surface term therefore also vanishes due to the periodicity condition on the integral boundaries. 
With the surface term vanishing and the integrand of the second integral always being positive, the expression \eqref{eq:av_back_force} is always smaller than or equal to zero.
The equality holds for a constant $\W$ corresponding to the case of a zero or constant $V(\phi)$.
We have therefore shown that any viable $V(\phi)$ leading to a sensible $\W$  reduces the fluctuations in $\phi$.

\paragraph{Validity of the classical fields approximation}

Since the Heisenberg and Hamiltonian equations of motion are identical for conjugate fields, our discussion is reduced to the initial state. 
From the validity of the classical fields approximation for the thermal initial state directly follows that the initial density fluctuations are Gaussian, which is an important prerequisite of our proposed mechanism for the emergence of Gaussian phase correlations.

Using the classical fields approximation for a thermal state corresponds to making two approximations: 
Firstly, the quantum (zero temperature) fluctuations are neglected. 
Secondly, the Bose-Einstein distribution for the occupation of the eigenmodes is replaced by the Rayleigh-Jeans distribution. 
Both approximations are good when the thermal occupation of the eigenmodes under consideration is sufficiently large.
Expressed as an equation the criteria read
\begin{equation}
	\frac{1}{\mathrm{e}^{\beta\epsilon_n} - 1} + \frac{1}{2} \approx \frac{1}{\beta\epsilon_n}, \label{eq:classical_crit}
\end{equation}
which is fulfilled for small values of $\beta\epsilon_n$.
Here $\beta = (k_\mathrm{B}T)^{-1}$ and $\epsilon_n$ are the mode energies.
Clearly the fulfillment of \cref{eq:classical_crit} depends on the mode-energy under consideration.
For the discussion we will therefore consider the highest mode that can still be resolved in experimental measurements.

For simplicity, we will use the quadratic approximation~\cite{Whitlock2003}
\begin{equation}
H = \int \mathrm{d} z \left[  \g(z) \, \delta \rho ^ { 2 } (z)  + \frac { \hbar ^ { 2 } \n(z)} { 4 m } \left( \frac { \partial  \varphi (z) } { \partial z } \right) ^ { 2 }
+ \hbar J\, \n(z)  \varphi^2 (z)  \right] \label{eq:H_quadratic}
\end{equation}
of \cref{eq:H_SG} for our discussion.
\Cref{eq:H_quadratic} is a good approximation for large phase-locking and, trivially, for zero tunneling.
Further assuming a homogeneous system, the eigenmodes are given by sine and/or cosine modes, depending on the boundary conditions.
The eigenenergy for the wavenumber $k$ is given by
\begin{equation}
	\epsilon = \hbar c \ \sqrt{k^2 + \frac{1}{l_J^2}}.
\end{equation}
Here $l_J$ is the healing length of the relative phase, it can be calculated as 
\begin{equation}
	l_J = \sqrt{\frac{\hbar}{4 m J}}.
\end{equation}
A sensible upper bound for $k$ can be derived from the spatial resolution of the phase measurement, which in turn can be bounded by the pixel size, $\Delta z_\mathrm{pix} = 2${\um}, of the used absorption imaging system.
We therefore use $k = 2\pi/\Delta z_\mathrm{pix}$ together with typical values for the temperature ($T = 70\, \mathrm{nK}$) and speed of sound ($c = 1.8\,\mathrm{mm/s}$).
For the tunneling rate $J$, we use the highest experimental value leading to the lowest $l_J = 2${\um}.
With this values the left and right hand side of \cref{eq:classical_crit} differ by only 3\%, justifying the use of the classical field approximation.  
Even though our argument was demonstrated using the quadratic approximation, we expect it to remain valid for the sine-Gordon model since the higher modes, where quantum effects might be relevant, are not affected strongly by the tunneling term.

\paragraph{The role of spatial mixing of correlations.}
\phantomsection
\label{sec:supp_info_ps_mix}

As found in earlier works, spatial mixing can only lead to the emergence of Gaussian correlations under two broadly applicable yet not generally valid conditions: 
The clustering of initial correlations~\cite{CramerEisert,Sotiriadis_2014,Gluza2016} and dynamical delocalization~\cite{Gluza2016,Sotiriadis_2017,GluzaEisertFarrelly,MurthySrednicki}.
The requirement of dynamical delocalization means that under the time evolution an initially localized disturbance will spread through the entire system and its amplitude will eventually decay.
It is easy to see that this condition is not satisfied by the dynamics of the homogeneous Luttinger liquid model~\cite{Sotiriadis_LL,Sotiriadis_2017}. 
Due to the phonon dispersion being linear, an initially localized wave packet does not spread but merely splits into a left- and a right-moving component that travel without dispersion, retaining in this way to large extent a memory of the initial correlations.
In the case of the non-homogeneous Luttinger liquid model (e.g.\ harmonic trap) the dynamics is not non-dispersive anymore, but still not delocalizing~\cite{Gluzainprep}.

To understand the effect of spatial mixing, let us consider the time evolution of the phase field $\varphi$ in coordinate space. 
For a general quadratic Hamiltonian, the Heisenberg equations of motion are linear and their solution for arbitrary initial conditions can be expressed as 
\begin{align*}
\varphi(z,t) & = \int {\rm d} z' \left(\,G_\varphi (z-z',t) \, \varphi(z',0) + G_{\delta\rho} (z-z',t) \,  \delta\rho(z',0) \, \right)
\end{align*}
where $G_\varphi $ and $G_{\delta\rho}$ are propagators of the quantum fields. 
The delocalization condition is satisfied roughly when the propagators spread in space so that their magnitude decays with time everywhere. 
From the above formula and based on the multi-linearity of correlation functions, we can easily express the dynamics of phase correlations $G_{\mathrm{con}}^{(N)}({\boldsymbol{z}},t)$ as spatial convolutions of the initial correlations of the fields $\varphi$ and $\delta\rho$. 
Then, assuming the validity of the delocalization condition together with the condition of initial clustering, it is possible to show that connected correlations of order higher than two decay with time. 
Given that correlation functions provide a complete characterization of the system's state, this proves that a thermodynamically large system will be locally described at large times by a Gaussian state. 
Moreover, finite size effects are bounded by the presence of a maximum bound in the speed of information propagation, which allows for a demonstration of this effect in a finite system.

In the case of the Luttinger liquid model dynamics, the Hamiltonian is given by \cref{eq:H_SG} with $J=0$.
In the homogeneous case, the Heisenberg equation of motion for the phase field $\varphi(x,t)$ is nothing but the 1D wave equation, whose solution in the infinite system is given by d'Alembert's formula
\begin{align*}
\varphi(z,t) & =\frac{1}{2}\left(\varphi(z+ct,0)+\varphi(z-ct,0)\right)-\frac{\g}{\hbar c}\int_{z-ct}^{z+ct}{\rm d}z'\delta\rho(z',0).
\end{align*}
Clearly, due to the absence of dispersion, the propagator $G_\varphi $ is localized at the edges of the light-cone $z'=z\pm ct$ and $G_{\delta\rho}$ is a simple step function that is a constant, independently of $t$, in the interior of the light-cone.
As a result, the dynamics of correlations $G_{\mathrm{con}}^{(N)}({\boldsymbol{z}},t)$ can be expressed as a relatively simple combination of initial phase correlations at the edges of the past light cone and initial density correlation (Gaussian in our model) from within it. 
Initial clustering still means that connected correlations show some decrease with time, due to the fact that correlations between distant points vanish.
However, this restricted mechanism of spatial mixing can only account for a limited decrease of the non-Gaussian correlations~\cite{Gluzainprep} as also apparent in \cref{fig:scan5100theowwodens}.

A variety of perturbations of different origin can lead to deviations from the linearity of the spectrum.
Such perturbations are for example the non-linearity of the phononic dispersion in the full Bogoliubov approximation~\cite{Mora2003} seen at the healing length scale, the presence of an inhomogeneous background density (e.g.\ harmonic trap) or phononic interactions. 
A careful analysis of the effects of such perturbations on the dynamics will be published in \rcite{Gluzainprep}. 
In particular, we find that in presence of a harmonic trap, the dynamics still is not delocalizing.
Also, the weak non-linearity in the phononic dispersion when considering the full Bogoliubov approximation instead of Luttinger liquid has only minor effects.
This can be understood from the fact that the healing length is about ten times smaller than the spatial resolution.
In the end, from observing the recurrences in the box-like trap, we know that the spectrum in this case is approximately linear despite the perturbations.